\documentclass[pra,twocolumn,nofootinbib,superscriptaddress,showpacs]{revtex4-1}
\usepackage{color,amsthm,amsmath,amsfonts,graphicx,bm}
\usepackage{bm}
\usepackage{bm,eufrak}
\usepackage[export]{adjustbox}
\usepackage{amsfonts}
\usepackage{amssymb}
\usepackage{amsmath,mathrsfs}
\usepackage{epsfig}
\usepackage{graphicx}
\usepackage{enumerate}
\usepackage{multirow}
\usepackage{verbatim}
\usepackage{subfigure}
\usepackage{bbold}
\usepackage{soul}
\usepackage{scrextend}
\usepackage{tikz}
\usepackage{comment}
\allowdisplaybreaks

\newcommand{\tr}{\mathrm{tr}}

\newcommand{\mr}[1]{\mathrm{#1}}
\newcommand{\dg}{\dagger}

\newcounter{notes}
\stepcounter{notes}

\DeclareFontFamily{OT1}{pzc}{}
\DeclareFontShape{OT1}{pzc}{m}{it}{<-> s * [1.10] pzcmi7t}{}
\DeclareMathAlphabet{\mathpzc}{OT1}{pzc}{m}{it}

\newtheorem{definition}{Definition}
\newtheorem{statement}{Statement}

\begin{document}

\title{Local Integrals of Motion for Topologically Ordered Many-Body Localized Systems}

\author{Thorsten B. Wahl}
\affiliation{DAMTP, University of Cambridge, Wilberforce Road, Cambridge, CB3 0WA, United Kingdom}
\author{Benjamin B\'eri}
\affiliation{DAMTP, University of Cambridge, Wilberforce Road, Cambridge, CB3 0WA, United Kingdom}
\affiliation{T.C.M. Group, Cavendish Laboratory, University of Cambridge, J.J. Thomson Avenue, Cambridge, CB3 0HE, United Kingdom}

\begin{abstract}
Many-body localized (MBL) systems are often described using their local integrals of motion, which, for spin systems, are commonly assumed to be a local unitary transform of the set 
of on-site spin-$z$ operators. We show that this assumption cannot hold for topologically ordered MBL systems. Using a suitable definition to capture such systems in any spatial dimension, we demonstrate a number of features, including that MBL topological order, if present: 
(i) is the same for all eigenstates; (ii) is robust in character against any perturbation preserving MBL; (iii) implies that on topologically nontrivial manifolds a complete set of integrals of motion must include nonlocal ones in the form of local-unitary-dressed noncontractible Wilson loops. Our approach is well suited for tensor-network methods, and is expected to allow these to resolve highly-excited finite-size-split topological eigenspaces despite their overlap in energy. We illustrate our approach on the disordered Kitaev chain, toric code, and X-cube model. 
\end{abstract}

\maketitle

\section{Introduction} 
Systems displaying many-body localization~\cite{Fleishman1980,gornyi2005interacting,*basko2006metal} (MBL) violate the eigenstate thermalization hypothesis~\cite{1984Peres,*deutsch1991quantum,*srednicki1994chaos,*Rigol:2008bh} and therefore do not thermalize. 
(See Refs.~\onlinecite{NandkishoreHuse_review,*AltmanReview,*Abanin2017,*Alet2017,ImbrieLIOMreview2017} for some recent reviews.)
MBL occurs in strongly disordered interacting lattice systems. 
Recent analytical and numerical work has put the effect on firm theoretical footing in one dimension~\cite{gornyi2005interacting,*basko2006metal,znidaric2008many,*pal2010mb,*Bardarson2012,*kjall2014many,
*imbrie2016many,Khemani2016MPS,*Pollmann2016TNS,Wahl2017PRX}, whereas the existence of MBL in higher dimensions is still debated~\cite{deRoeck2017Stability,Altman2018stability}. However, MBL has been observed experimentally both in one-~\cite{Schreiber842,*Lukin2018,*Smith_MBL,*Roushan2017} and two-dimensional 
systems~\cite{Choi1547,*bordia2017quasiperiodic2D,*2D_quantum_bath}. This might be due to extremely long relaxation times~\cite{chandran2016higherD}, which are unobservable in the experiments. Numerical simulations are consistent with two-dimensional MBL-like behavior~\cite{2DMBL,*Kennes2018,*Alet2D,*Kshetrimayum2019,*Doggen20}.

All eigenstates of MBL systems are area-law entangled~\cite{2013Bauer_Nayak,Friesdorf2015}. 
This makes MBL compatible with the scenario where all eigenstates are topologically ordered~\cite{Huse2013LPQO,2013Bauer_Nayak,bahri2015localization,2015Potter,Parameswaran2018,Thorsten,1DSPTMBL,2DSPTMBL,topMBL_phase_diagram,*Kuno2019}. 
In one-dimensional systems, nontrivial topology requires the presence of certain symmetries. In higher dimensions, however, topological states can exist without symmetries and have fractionalized quasiparticles (anyons) and spectral degeneracies dependent only on the topology of the system's configuration space (the two features are interlinked~\cite{Oshikawa06}).
It is these states that we call here topologically ordered, while we call topological states requiring symmetries for their existence symmetry-protected topological (SPT) states~\cite{Wen_review}. 
Topologically ordered MBL systems have been suggested to provide improved protection of quantum information against perturbations compared to their clean counterparts~\cite{WoottonPRL2011,*StarkPRL2011,2013Bauer_Nayak}.

Fully MBL (FMBL) systems can be described in terms of \textit{local integrals of motion (LIOMs)}, which are exponentially localized operators which commute with each other and the Hamiltonian~\cite{serbyn2013local,Huse_MBL_phenom_14,chandran2015constructing,ros2015integrals,Inglis_PRL2016,
Rademaker2016LIOM,Monthus2016,ImbrieLIOMreview2017,Goihl2018,*Abi2017,
Abi2019}. LIOMs are commonly assumed to form a complete set  arising as a local unitary transform of the set of on-site spin-$z$ operators.
This tacitly assumes the absence of topological order: it implies that FMBL eigenstates arise from product states of this spin-$z$ basis via local unitary transformation, which guarantees~\cite{Bravyi2006,Chen_Gu,*HastingsPRL2011} that they are topologically trivial. 

Topologically ordered FMBL systems therefore require a more general notion of LIOMs.  
The key goals of this paper are to (i) develop such a ``\emph{topological LIOM (tLIOM)}" notion and thereby provide a precise definition of FMBL with topological order; to investigate (ii) what properties of topological FMBL phases follow from this definition, and (iii) how tLIOMs may be used to characterize such phases in numerical simulations. 
The concept of tLIOMs can be illuminated by placing LIOMs into the broader context of the \emph{stabilizer formalism}~\cite{NielsenChuang}.
From this perspective, one sees the on-site \mbox{spin-$z$} operators 
as just one choice of local stabilizers, namely those of product states in this spin-$z$ basis. 
Topological LIOMs must then correspond to a different set, namely the local stabilizers in the commuting projector limits of topological phases. 
These tLIOMs are closely related to the approach of Ref.~\onlinecite{2015Potter} to many-body localizability, where the use of commuting projector limits as reference points was first suggested and used to investigate topological MBL states without fractionalized quasiparticles (i.e., ``integer" topological states and SPTs).
However, since all non-chiral forms of Abelian topological order admit such commuting projector limit~\cite{Levin_Wen}, the tLIOMs capture all Abelian non-chiral FMBL eigenstate topological orders. 
The scope is not restricted by dimensionality; it even includes putative FMBL cousins of recently introduced fractonic phases~\cite{Haah2011,X-cube,Shirley2018,Fractons}.
Furthermore, tLIOMs lend themselves to be combined with tensor-network methods thus far used for non-topological FMBL systems~\cite{Khemani2016MPS,*Pollmann2016TNS,Wahl2017PRX,2DMBL,*Kennes2018,*Kshetrimayum2019}. 
This tLIOM---tensor-network combination is the perspective through which we shall seek new avenues for the numerical characterization of topological FMBL systems.

\section{Non-topological FMBL}  
\begin{figure}
\includegraphics[width=0.45\textwidth]{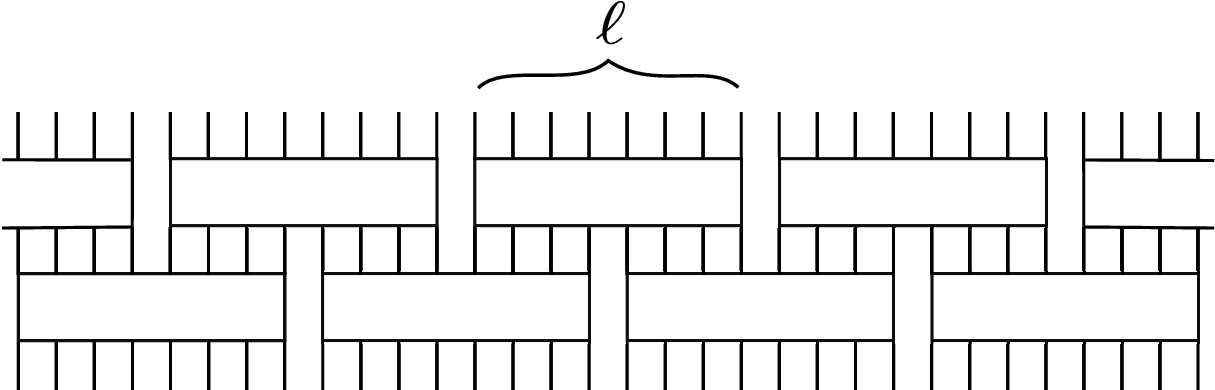}
  \caption{A  one-dimensional depth-two quantum circuit with length-$\ell$ unitaries (denoted as boxes). In $d$-dimensional systems, $\ell$ is allowed to grow sublinearly with $N^{1/d}$~\cite{Wahl2017PRX}.}    
\label{figure}
\end{figure}

For concreteness we consider an $N$-site spin-$1/2$ (i.e., qubit)  system on a $d$-dimensional square lattice. The extension to higher-spin systems and fermionic systems is straightforward. 
The FMBL phase is defined by a complete set of LIOMs $\tau_i^z$ ($i=1,\ldots,N$) which commute with the Hamiltonian $H$ and each other,
\begin{align}
[H,\tau_i^z] = [\tau_i^z, \tau_j^z] = 0, \label{eq:commutation}
\end{align}
and are exponentially localized, i.e., their non-trivial matrix elements decay exponentially with distance from site $i$. The corresponding decay length, the \textit{localization length} $\xi_i$, must fulfill $\xi_i/N^{1/d}\rightarrow 0$ for all $i$ in the thermodynamic limit $N\rightarrow \infty$. 
The $\tau_i^z$ can be constructed from a unitary $U$ which diagonalizes the Hamiltonian as $\tau_i^z = U \sigma_i^z U^\dagger$, where $\sigma_i^z$ is the third Pauli operator acting on site $i$. 
A key feature that makes nontopological FMBL systems special, and computationally tractable~\cite{Khemani2016MPS,*Pollmann2016TNS,Wahl2017PRX,2DMBL}, is that $U$ is a
\emph{local unitary}, i.e., efficiently approximable by a constant-depth quantum circuit of  unitary gates with length  (i.e., linear size) $\ell$ sublinear in $N^{1/d}$~\cite{Wahl2017PRX}. [The error scales as $\exp(-\ell/\xi_\text{max})$ where $\xi_\text{max}$ is the largest $\xi_i$.]
Fig.~\ref{figure} shows a depth-two example. 

The $\tau_i^z$'s are also known as \textit{l-bits} (l=localized)~\cite{Huse_MBL_phenom_14}; they label, due to Eq.~\eqref{eq:commutation}, all eigenstates $|\psi_{l_1 \ldots l_N}\rangle$ of $H$,
\begin{align}\label{eq:ptolbit}
\tau_i^z |\psi_{l_1 \ldots l_N}\rangle = l_i |\psi_{l_1 \ldots l_N}\rangle,\quad l_i = \pm 1. 
\end{align}
The $\sigma_i^z$ operators are called \textit{p-bits} (p=physical)~\cite{Huse_MBL_phenom_14}. Eq.~\eqref{eq:ptolbit} implies $|\psi_{l_1 l_2 \ldots l_N}\rangle = U |l_1 l_2 \ldots l_N\rangle$ in terms of the p-bit product states $|l_1 l_2 \ldots l_N\rangle$. 
For $d > 1$, the LIOM description might not apply exactly due to delocalization on extremely long time scales~\cite{deRoeck2017Stability,chandran2016higherD}. However, on experimental time scales, the description in terms of LIOMs seems to be appropriate~\cite{2DMBL}, which is what we restrict ourselves to in the following.

\section{Topological FMBL phases} 

In the topological FMBL case it is conjectured that all eigenstates display
topological order~\cite{Huse2013LPQO,2013Bauer_Nayak,Parameswaran2018}. However, even if only \textit{one} eigenstate $|\psi_{l_1 l_2 \ldots l_N}\rangle$ is topologically non-trivial, there exists no local unitary $U$ such that $|\psi_{l_1 l_2 \ldots l_N}\rangle = U |l_1 l_2 \ldots l_N\rangle$
~\cite{Bravyi2006,Chen_Gu,*HastingsPRL2011}. 
Hence, if a LIOM description applies, it is not of the form $\tau_i^z = U \sigma_i^z U^\dg$. 
We must therefore use a more general notion of (t)LIOMs appropriate also for topological FMBL systems. 

We begin by explaining how p-bits and the corresponding product eigenstates fit in the broader context of \textit{local stabilizer codes}~\cite{NielsenChuang} (see also Ref.~\onlinecite{2015Potter} for a closely related approach to MBL).
For concreteness, we present the idea for $N$-site qubit systems, however the scope of stabilizer codes, and hence our construction, is much more general. 
A local stabilizer code may be thought of in terms of a stabilizer Hamiltonian
\begin{align}
H_\mr{sc} = -\sum_{i=1}^N S_i, \label{eq:Ham_sc}
\end{align}
where the local stabilizers $S_i = S_i^\dg$ are Pauli strings with local support around site $i$, and $[S_i,S_j] = 0$. (Thus, up to a constant, $H_\mr{sc}$ is a commuting-projector Hamiltonian.) On a topologically trivial manifold, $S_i$ are independent ($\prod_i S_i^{x_i}=1$ only if $x_i=0,\ \forall\ i;\  x_i\in\{0,1\}$), hence their eigenvalues $s_i=\pm1$ label a complete orthonormal basis. We refer to $S_i$ as \textit{s-bits}.
The p-bits and their product eigenstates correspond to $S_i = \sigma_i^z$ and the eigenstates of $H_\mr{sc}$, respectively. 
The scope of stabilizer codes, however, is much wider (and not limited to qubits): they capture all non-chiral topological orders~\cite{Levin_Wen}, from the toric code~\cite{toric_code} to fracton models~\cite{Haah2011,X-cube,Shirley2018,Fractons}, as well as fermionic systems such as the Kitaev chain~\cite{Kitaev_chain} or various other Majorana fermion codes~\cite{bravyi2010majorana}. 
We propose the following definition of FMBL and (t)LIOMs to capture Abelian topological order on topologically trivial manifolds:
\begin{definition}
\label{def:S_FMBL}
Let $S_i$, $i = 1, \ldots, N$ be a complete set of local stabilizers on a topologically trivial manifold $\mathcal{M}$. 
We call a local Hamiltonian $H$ on $\mathcal{M}$ FMBL, if there exists a local unitary $\tilde U$ such that $T_i = \tilde U S_i \tilde U^\dg$,
\begin{align}\label{eq:tFMBL}
[H, T_i] = [T_i, T_j] = 0,
\end{align}
and the same property holds in an $\epsilon > 0$ open environment around $H$ [i.e., for local perturbations $\delta H$ of strength $\| \delta H \| / \| H \| < \epsilon$ where $\|\cdot \|$ is a suitable norm], where for $N$ sufficiently large, $\epsilon$ does not depend on $N$ [i.e., $\epsilon$ is nonzero in the thermodynamic limit]. We call the complete set $T_i$ tLIOMs (or topological l-bits) if $S_i$ are the stabilizers of topologically ordered states.
\end{definition}

We shall come back to discussing various aspects of our definition, including observations for topologically nontrivial $\mathcal{M}$ and non-Abelian topological order. For now, we note that Definition~\ref{def:S_FMBL} implies, firstly,  $|\psi_{\{s_i\}}\rangle=\tilde{U}|\{s_i\}\rangle$ where $|\psi_{\{s_i\}}\rangle$ and $|\{s_i\}\rangle$ are respective eigenstates of the FMBL Hamiltonian $H$ and $H_\text{sc}$.
Secondly, up to an additive constant, and with $c_{ijk\ldots} \in \mathbb{R}$,
\begin{align}
H =  \sum_i c_i T_i + \sum_{i < j} c_{ij} T_i T_j  + \sum_{i<j<k} \!\! c_{ijk} T_i T_j T_k + \ldots, \label{eq:expansion}
\end{align}
where for a local Hamiltonian and for describing the dynamics on experimental time scales, we can assume that $|c_{ijk\ldots}|$ decay exponentially with the largest distance between the locations $i,j,k,\ldots$ in any dimension $d$. 

We next establish a number of features that follow from Definition~\ref{def:S_FMBL}, focusing again on qubit systems for concreteness. 
We first remind of the following result~\cite{Bravyi2006,Chen_Gu,*HastingsPRL2011}: $|\psi_1\rangle$ and $|\psi_2\rangle$ have the same topological order if $|\psi_2\rangle=U|\psi_1\rangle$ and $U$ is a local unitary. Together with Definition~\ref{def:S_FMBL}, this means that many features directly carry over from the commuting projector limit $H_\text{sc}$ to the FMBL phase. For instance, due to $|\psi_{\{s_i\}}\rangle=\tilde{U}|\{s_i\}\rangle$:
\begin{statement}
\label{st:psi_sim_s}
The FMBL eigenstates have the same topological order as those of $H_\mr{sc}$.
\end{statement}
Furthermore, since any set of the stabilizers $S_i$ can be flipped by a suitable Pauli string~\cite{NielsenChuang}, and since any Pauli string is a local unitary, every eigenstate of $H_\mr{sc}$ has the same topological order. Hence:
\begin{statement}
\label{st:psi_all_sameTO}
All eigenstates of topological FMBL systems display the same topological order. 
\end{statement}

Definition~\ref{def:S_FMBL} requires that if the system is FMBL, it should also be FMBL after having applied a local perturbation (e.g., a translation-invariant nearest-neighbor coupling term) of relative strength $\epsilon$, and this should hold for $\epsilon$ sufficently small but nonzero even in the thermodynamic limit: otherwise we would be at a phase transition point. 
The requirement of robustness against perturbations in a nonzero environment around the Hamiltonian $H$ thus amounts to describing a localized \textit{phase}. 

Conversely, Definition~\ref{def:S_FMBL} excludes systems with (t)LIOMs that easily delocalize.  
An example of such systems is $H_\mr{sc}$ itself [which satisfies Eq.~\eqref{eq:tFMBL} with $\tilde{U}=1$], as illustrated by $S_i=\sigma_i^z$ and $H(\delta t)=H_\mr{sc}+\delta t\sum_i(\sigma_i^x \sigma_{i+1}^x+\sigma_i^y \sigma_{i+1}^y)$ in $d=1$: for any $\delta t\neq 0$, Jordan-Wigner transformation reveals the integrals of motion as plane-wave-operators, hence not related to $S_i$ by a local unitary.
For a local stabilizer code with topological order, the stabilizers with $s_i=-1$ indicate anyon locations; in the FMBL phase these translate to the support of the corresponding $T_i$. By Definition~\ref{def:S_FMBL}, this support does not easily delocalize: we find anyon localization.

In Definition~\ref{def:S_FMBL} we specialized to a topologically trivial $\mathcal{M}$ and Abelian topological order so that %
eigenstates are fully characterized by the \emph{local} s-bit strings $\{s_i\}$. For a topologically nontrivial $\mathcal{M}$ 
and/or non-Abelian topological order, the set of local $S_i$ is \emph{not} complete; $\{s_i\}$ label  subspaces of degeneracy depending on the topology of $\mathcal{M}$ and/or the anyon fusion~\cite{Kitaev2006}. A complete set  includes \emph{nonlocal} $S^\text{nl}_i$ required to resolve these degeneracies. 
In the non-Abelian case, due to the nonzero density of anyons in generic eigenstates, the number of $S^\text{nl}_i$ is extensive; tLIOMs give a highly incomplete characterization~\cite{vass201612}. Hence we focus on the Abelian case. There, $S^\text{nl}_i$ are noncontractible Wilson loops (e.g., certain Pauli strings for qubits) on $\mathcal{M}$. We can thus complete the set of FMBL integrals of motion by $T^\text{nl}_i = \tilde U S^\text{nl}_i \tilde U^\dg$. Therefore:
\begin{statement}
\label{st:psi_top_nontriv}
On a topologically nontrivial $\mathcal{M}$, the complete set of FMBL integrals of motion must include $T^\mr{nl}_i = \tilde U S^\mr{nl}_i \tilde U^\dg$ where $S^\mr{nl}_i$ are noncontractible Wilson loops resolving the eigenspace degeneracies of $H_\mr{sc}$.  
\end{statement}
\noindent These $T^\mr{nl}_i$ give an operational definition of the fattened Wilson loops in Ref.~\onlinecite{2013Bauer_Nayak}. Aiming at $\{T_i\}$ via $\{S_i\}$ in a tensor-network calculation~\cite{Khemani2016MPS,*Pollmann2016TNS,Wahl2017PRX,2DMBL}, one may in principle resolve highly-excited topological multiplets despite the huge density of states. 

We next introduce an FMBL notion of topological equivalence:
\begin{definition}
\label{def:FMBL_top_equiv}
Two FMBL Hamiltonians $H_0$ and $H_1$ are in the same topological phase if and only if for all sufficiently large $N$ there exists a continuous parameterization $H(\lambda)$, $\lambda \in [0,1]$, with $H(0) = H_0$ and $H(1) = H_1$, such that $H(\lambda)$ is FMBL for all $\lambda \in [0,1]$ and it converges to a continuous parameterization as $N\rightarrow \infty$.
\end{definition}
\noindent In other words, one cannot connect topologically inequivalent FMBL Hamiltonians without delocalizing the system along the way. Furthermore, 
\begin{statement}
\label{st:FMBLvsEigenstate}
Two FMBL Hamiltonians $H_0$ and $H_1$ are in the same topological phase if and only if their eigenstate topological order is the same. 
\end{statement}
We demonstrate this in the scope of Definition~\ref{def:S_FMBL}. 
We first show that the same eigenstate topological order implies the same FMBL topological phase. 
If the eigenstate topological order is the same, then the sets $\{S_{i,0}\}_{i=1}^N$ and $\{S_{i,1}\}_{i=1}^N$ corresponding to Hamiltonians $H_0$ and $H_1$, respectively, are mapped by a local unitary $U_S$ (with degree of locality linked to those of the eigenstate-mapping unitaries): $S_{i,1} = U_S S_{i,0} U_S^\dg$ for all $i$. 
We consider $H_{\alpha}$ (with $\alpha = 0,1$) expanded according to Eq.~\eqref{eq:expansion}
with $T_{i,\alpha} =\tilde U_\alpha S_{i,\alpha} \tilde {U}_\alpha^\dg$ and coefficients $c_{ijk\ldots,\alpha}$. 
We define $c_{ijk\ldots}(\lambda) = (1-\lambda) c_{ijk\ldots,0} + \lambda c_{ijk\ldots,1}$. 
We also define a local unitary $\tilde U(\lambda)$ such that $\tilde U(0) = \tilde U_0$ and $\tilde U(1) = \tilde U_1 U_S$ and $\tilde U(\lambda)$ a continuous function of $\lambda$~\cite{Chen_Gu}.
$H(\lambda)$ defined as in Eq.~\eqref{eq:expansion} 
gives a continuous path connecting $H_0$ and $H_1$ preserving FMBL (Definition~\ref{def:S_FMBL}) for all $\lambda \in [0,1]$.

To show the converse, we use the fact that $H(\lambda)$ fulfills Definition~\ref{def:S_FMBL} for all $\lambda \in [0,1]$. Hence, FMBL is preserved in an $\epsilon(\lambda)$ environment around $H(\lambda)$. 
We next observe that the eigenstate topological order is the same across this  $\epsilon(\lambda)$ environment: otherwise, since topological order cannot be changed continuously~\cite{Kitaev2006} (as embodied by the discreteness of the topological equivalence classes of stabilizers~\cite{Levin_Wen}), the environment would have to contain points where the spectrum of the system has degeneracies such that the local perturbation $\delta H$ in    Definition~\ref{def:S_FMBL} can switch the topological order. 
However, Definition~\ref{def:S_FMBL} also requires that (t)LIOMs stay local, hence degenerate states that can be coupled by $\delta H$ must differ at most in the action of a locally supported operator. (In other words, they must form dilute, well-isolated, small resonant clusters~\cite{Potter:2015ab}.)
The corresponding unitary rotations in the degeneracy spaces amount to a local unitary update of $\tilde{U}$, hence topological order cannot be switched. 
Using this constancy of the topological order within the $\epsilon(\lambda)$ environments, the rest of our demonstration is straightforward: the continuity of $H(\lambda)$, together with the fact that $I_\lambda=[0,1]$ is compact and connected, implies that $H(I_\lambda)$ is compact and connected. Its open cover consisting of the $\epsilon(\lambda)$ environments thus has a finite subcover which we can choose such that successive (in $\lambda$) environments overlap with each other. 
[The $N$-independence of $\epsilon(\lambda)$ ensures that the number of $\epsilon$-environments in this subcover is $N$-independent.]
Hence eigenstate topological order is the same on the entire path $H(I_\lambda)$: $H_0$ and $H_1$ must have the same topological order.

An implication of Statement~\ref{st:FMBLvsEigenstate}  is that the (t)LIOMs along the path $I_\lambda$ can be written as $T_j(\lambda)=\tilde{U}(\lambda)S_j\tilde{U}^\dagger(\lambda)$ such that $\tilde{U}(\lambda)$ remains a local unitary for any $\lambda$. 
Going along $I_\lambda$, one encounters resonances at certain values of $\lambda$. While, as we noted in our demonstration, FMBL requires these to form small resonant clusters, the set of points where such resonances occur becomes increasingly dense in $I_\lambda$ as $N\rightarrow \infty$ due to the increasing number of possible spatial locations for these clusters. 
One might thus wonder why the corresponding extensive number of local unitary updates applied successively as we go along $I_\lambda$ cannot result in $\tilde{U}(\lambda)$ ceasing to be a local unitary. 
An intuitive reasoning for this is as follows: 
Firstly, the locality of a resonant cluster implies that, upon crossing the corresponding value in $\lambda$, the matrix $\tilde{U}(\lambda)$ is multiplied by is not merely a local unitary, but (to exponential accuracy) a locally-supported gate with size set by that of the resonant cluster, the localization lengths $\xi_i$ of the corresponding (t)LIOMs and the decay length of the couplings in Eq.~\eqref{eq:expansion}. 
The locality of FMBL physics implies that in any fixed-size spatial region, the number of such resonance clusters occurring as one crosses $I_\lambda$ becomes $N$-independent for large $N$. 
Moreover, due to the randomness inherent to FMBL systems, the corresponding local gates appear in random locations. 
These considerations lead to a circuit with $N$-independent depth and gate-length sublinear in $N^{1/d}$, hence the cumulative action corresponding to these resonances is a local unitary. 

\section{Topological LIOMs: Examples}

Next, we illustrate our tLIOMs on three examples, including qubits and fermions in $d=1,2,3$. We point out which non-local integrals of motion emerge for those systems and how they give rise to approximate degeneracies arbitrarily high up in the spectrum. For the disordered toric code ($d = 2$), we suggest a way of resolving such almost degenerate eigenstates despite the significantly smaller average level spacing. 

\subsection{Disordered Kitaev chain ($d=1$)}

This is the spinless-fermion Hamiltonian on an $N$-site open chain~\cite{Kitaev_chain}
\begin{equation}
H_\mr{K} = \frac{1}{4} \sum_{n = 1}^{N-1} t_n \left(a_n a_{n+1} + a_n a_{n+1}^\dg + h.c.\right),
\end{equation}
where $a_n$ annihilates a fermion at site $n$ and $t_n$ is Gaussian-random with zero  mean and unit variance.
Introducing the Majorana operators 
\begin{equation}
\gamma_{2n-1} = \frac{1}{2} (a_n + a_n^\dg), \quad
\gamma_{2n} = \frac{1}{2} (-i a_n + i a_n^\dg),
\end{equation}
the Hamiltonian can be rewritten as
\begin{equation}\label{eq:Kitaev_Majorana}
H_\mr{K} = i \sum_{n=1}^{N-1} t_n \gamma_{2n} \gamma_{2n+1}.
\end{equation}
Although it is a (non-interacting) commuting projector Hamiltonian ($S_n = i \gamma_{2n} \gamma_{2n+1}$), it fulfills Definition~\ref{def:S_FMBL} because the localization of all eigenstates is stable as interactions are introduced~\cite{Huse2013LPQO}.
It is also topological: 
The s-bits $S_n$ ($n = 1,\ldots, N-1$), completed by $S_N^\text{nl} = i \gamma_{2N} \gamma_1$, cannot be connected to local fermion-product-state p-bits  $(-1)^{a_n^\dagger a_n}$ by any fermion-parity conserving local unitary~\cite{Turner2011}. (The same holds for a closed chain where  $S_N$ is also local.) Here, fermion-parity is a protecting symmetry, although often taken as given in which case Eq.~\eqref{eq:Kitaev_Majorana} is considered topologically ordered. 
For Eq.~\eqref{eq:Kitaev_Majorana}, the tLIOMs are $T_n=S_n$; upon adding weak disorder (e.g., $i \sum_{n=1}^N \mu_n  \gamma_{2n-1} \gamma_{2n}$ with zero-mean Gaussian-random $\mu_n$ of much smaller than unit variance) they become $T_n = \tilde U S_n\tilde U^\dg$, where $\tilde U$ is a parity-conserving local unitary. In addition to $T_n$, the nonlocal $T_N^\text{nl} = \tilde U  S_N^\text{nl}\tilde U^\dg$ also appears in the expansion~\eqref{eq:expansion}, but with  magnitude exponentially suppressed in the linear system size $L$. We note that the topological phase characterized by the tLIOMs $T_n = \tilde U i \gamma_{2n} \gamma_{2n+1} \tilde U^\dg$ accounts for the missing $\mathbb{Z}_2$ index in the classification of fermionic one-dimensional topological MBL phases (with a symmetry) using only short-depth quantum circuits~\cite{1DSPTMBL}.

\subsection{Disordered Toric code ($d=2$)}

We consider qubits on the links of a square lattice on a torus and
\begin{align}\label{eq:toric_code}
H_\mr{tc} = - \sum_v J_v A_v - \sum_p K_p B_p,
\end{align}
where $A_v = \prod_{i \in v} \sigma^x_i$ and $B_p = \prod_{i \in p} \sigma^z_i$ act on vertices $v$ and plaquettes $p$ of the lattice, respectively~\cite{toric_code}.  The couplings $J_v$ and $K_p$ are again Gaussian-random with mean 0 and variance 1. Eq.~\eqref{eq:toric_code} is also a commuting projector Hamiltonian, but is expected to fall under Definition~\ref{def:S_FMBL} because upon adding weak disorder, such as a randomly fluctuating magnetic field $\sum_i h_i \sigma^z_i$, the Hamiltonian is believed to remain FMBL~\cite{2013Bauer_Nayak,Parameswaran2018}.  
The local s-bits are $\{A_v\}, \{B_p\}$; they are completed by, e.g., the noncontractible Wilson loops $S_{1,2}^\text{nl}=\mathcal{Z}_{1,2} = \prod_{i \in \mathcal{C}_{1,2}} \sigma_i^z$ on the two  generating cycles $\mathcal{C}_{1,2}$ of the torus. Hence, s-bit strings specify  topologically degenerate eigenspaces $|\{s_i\},z_1,z_2\rangle$ with $z_i=\pm1$ the eigenvalue of $\mathcal{Z}_i$. 
Upon adding weak disorder, tLIOMs become $\{\tilde U A_v \tilde U^\dg\}$, $\{\tilde U B_p \tilde U^\dg\}$, where again $\tilde U$ is a local unitary.  These are completed by the nonlocal integrals of motion, e.g.,  $\tilde U \mathcal{Z}_{i} \tilde U^\dg$; as before, for finite linear system size $L$ these also appear in the expansion~\eqref{eq:expansion}, but with exponentially suppressed coefficients. 
The corresponding $\propto\exp(-L)$ splitting of topological degeneracies is much larger than the level spacing $\propto\exp(-L^2)$. 
This makes detecting topological multiplets impossible in exact diagonalization~\cite{Huse2013LPQO,2013Bauer_Nayak,Parameswaran2018}. 
As we noted earlier, our framework can avoid this problem: it allows us to take advantage of the efficient approximation of $\tilde U$ by a quantum circuit $U_\mr{qc}$, and employ the methods of Refs.~\cite{Wahl2017PRX,2DMBL} to  numerically minimize $\sum_i \tr\left([H,T_{\mr{qc,}i}][H,T_{\mr{qc},i}]^\dg\right)$ with $T_{\mr{qc},i} = U_{\mr{qc}} S_i U_\mr{qc}^\dg$. 
Thus obtaining $U_{\mr{qc}}$ gives the approximation of the tLIOMs $T_{\mr{qc,}i}$, of their nonlocal completion $T^{\text{nl}}_{\mr{qc,}i}=U_{\mr{qc}} \mathcal{Z}_{i} U_\mr{qc}^\dg$, and of the topological multiplets $U_{\mr{qc}} |\{s_i\},z_1,z_2\rangle$. 
In addition to demonstrating these FMBL topological multiplets, the presence of FMBL topological order could be tested by comparing to minimizing with $\tau_{\mr{qc},i} = U_{\mr{qc}} \sigma_i^z U_\mr{qc}^\dg$ instead of $T_{\mr{qc},i}$.  If $T_{\mr{qc},i}$ perform significantly better, this would indicate that the system is in a topological FMBL phase. 
\subsection{Disordered X-cube model ($d=3$)}

Fractons are emergent excitations which either cannot move without creating additional fractons (at an energy cost) or can move only along certain directions~\cite{Haah2011,X-cube,Shirley2018,Fractons}. 
Here, we focus on the so-called X-cube model~\cite{X-cube} of fractons of the latter type. 
This model has qubits on the links of a cubic lattice on the 3-torus. The Hamiltonian is
\begin{align}
H_\mr{X} = - \sum_c u_c A_c - \sum_{v;\mu\in \{xy,xz,yz\}} K_{v}^{\mu} B_v^{\mu},
\end{align}
where $A_c = \prod_{i \in c} \sigma_i^x$, $B_v^{\mu} = \prod_{i \in v(\mu)} \sigma_i^z$ with $c$ denoting a cube and $v(\mu)$ the sites around vertex $v$ lying parallel to  plane $\mu = xy, xz, yz$. The couplings $u_c$ and $K_v^{\mu}$ are Gaussian-random with mean 0 and variance 1.  
This, again, is a commuting projector Hamiltonian, but may satisfy Definition~\ref{def:S_FMBL} as fracton models may become FMBL~\cite{Parameswaran2018}.
The local s-bits are $\{A_c\}, \{B_v^\mu\}$. On a 3-torus of linear size $L$, they are completed by $6L-3$ independent noncontractible commuting Wilson loops $S_{i}^\text{nl}$~\cite{Shirley2018}. The subextensive scaling of the number of these suggests that a tLIOM description may be useful, which proceeds analogously to the toric code case. 

\section{Conclusion and Outlook}

Defining topological FMBL phases using tLIOMs provides a transparent framework for establishing a number of features on the character and robustness of these phases. 
Firstly, it allowed us to show that (i) all eigenstates of topological MBL systems must have the same topological order (Statement~\ref{st:psi_all_sameTO}) and that (ii) this order must be the same throughout the topological FMBL phase (Statement~\ref{st:FMBLvsEigenstate}). The topological properties of all eigenstates are thus robust to small perturbations; changing them requires delocalization (see Ref.~\onlinecite{Sze20} for a numerical study). 
Features (i) and (ii) are shared~\cite{Thorsten,1DSPTMBL,2DSPTMBL} with FMBL SPT systems~\cite{bahri2015localization,2015Potter,topMBL_phase_diagram,*Kuno2019},
however unlike in those cases [which \emph{are} local-unitary related to on-site product states and hence allow conventional LIOMs 
to be used to establish (i) and (ii) for SPTs], establishing these results in the topologically ordered case relies on the tLIOM approach in an essential way. 
Furthermore, our approach also allowed us to capture features without SPT counterparts. In particular, we showed
(iii) how anyon localization (implied by Definition~\ref{def:S_FMBL}) and (iv) spectral degeneracy on topologically nontrivial manifolds (Statement~\ref{st:psi_top_nontriv}) follow from our framework. 
These results not only put certain thus far only heuristically established findings~\cite{WoottonPRL2011,*StarkPRL2011,2013Bauer_Nayak} on firm footing, but en route to (iv) we also demonstrated the existence, and provided the operational definition, of \emph{nonlocal} integrals of motion underlying the topological multiplets and complementing the tLIOMs to form a complete set. 
Besides providing a transparent theoretical picture, such an operational definition has practical significance: upon combining our tLIOMs with tensor-network approaches, it allows one, e.g., to numerically resolve topological multiplets at high energies, an objective hitherto considered infeasible %
due to the mean level spacing scaling faster to zero with system size than the multiplet splitting~\cite{Huse2013LPQO,2013Bauer_Nayak,bahri2015localization,Parameswaran2018}.

In closing, we mention a few examples of future directions where our tLIOM framework may find uses or generalizations. 
The l-bit description gave key insights into the phenomenology of FMBL systems, including into the dynamics of quantum information~\cite{Huse_MBL_phenom_14}. 
A natural question is: what new features may arise in tLIOM-Hamiltonians~\eqref{eq:expansion} due to topological order?
The suitability of our framework for tensor-network methods also opens the door for numerically addressing questions with at most heuristic answers thus far: under what conditions is FMBL  present e.g., in the disordered toric code and what level of improvement may FMBL provide in protecting the encoded quantum information? 
How do the conditions for FMBL depend on the type of Abelian topological order beyond the toric code?  
How do tLIOMs behave near the topological MBL transition? 
FMBL is often invoked for protecting driven (Floquet) phases from heating~\cite{harper2019floquet}; our framework may thus find applications in novel topologically ordered driven phases of matter. 
It would also be interesting to generalize our approach to symmetry-enriched topological  phases~\cite{Wen_review} in the FMBL regime. The quantum circuit formalism of Ref.~\onlinecite{2DSPTMBL} might be of particular relevance for this endeavor.

\acknowledgments
This research was supported by the ERC Starting Grant No. 678795 TopInSy.

\bibliography{biblioMBL}

\end{document}